# #Halal Culture on Instagram

**Yelena Mejova** [1], **Youcef Benkhedda** [2], **Khairani** [3]

[1]*Qatar Computing Research Institute, Hamad Bin Khalifa University, Doha, Qatar*
[2]*Ecole Nationale Supérieure d'Informatique, Oued Smar, Algeria*
[3]*Faculty of Economics and Business, Universitas Indonesia, Indonesia*

Correspondence*:
Yelena Mejova
yelenamejova@acm.org

## ABSTRACT

Halal is a notion that applies to both objects and actions, and means *permissible* according to Islamic law. It may be most often associated with food and the rules of selecting, slaughtering, and cooking animals. In the globalized world, halal can be found in street corners of New York and beauty shops of Manila. In this study we explore the cultural diversity of the concept, as revealed through social media, and specifically the way it is expressed by different populations around the world, and how it relates to their perception of (i) religious and (ii) governmental authority, and (iii) personal health. Here, we analyze two Instagram datasets, using Halal in Arabic (325,665 posts) and in English (1,004,445 posts), which provide a global view of major Muslim populations around the world. We find a great variety in the use of halal within Arabic, English, and Indonesian-speaking populations, with animal trade emphasized in first (making up 61% of the language's stream), food in second (80%), and cosmetics and supplements in third (70%). The commercialization of the term halal is a powerful signal of its detraction from its traditional roots. We find a complex social engagement around posts mentioning religious terms, such that when a food-related post is accompanied by a religious term, it on average gets more likes in English and Indonesian, but not in Arabic, indicating a potential shift out of its traditional moral framing.

## 1 INTRODUCTION

*Generation M*, as defined by Shelina Janmohamed, is the young generation of Muslims which embraces modernity, as well as its faith (Janmohamed, 2016). Currently, Muslims have the youngest median age (23 in 2010) of all major religious groups, seven years younger than the median age of non-Muslims (30). Pew Research Center projects Muslims will grow more than twice as fast as the overall world population between 2010 and 2050, likely surpassing Christians as the world's largest religious group[1]. As the generation of Muslim millenials asserts itself in the web space, it brings its culture to the world's social media stage, potentially redefining its traditions.

---

[1] http://www.pewresearch.org/fact-tank/2015/04/23/why-muslims-are-the-worlds-fastest-growing-religious-group/

An important part of Muslim daily life is the notion of *halal* – that which is *permissible* according to the Islamic law. The UN Food and Agriculture Organization (FAO) defines "halal" in the context of restrictions for foods of animal and plant origins, drinks, and food additives, which are not "unlawful according to Islamic Law"[2]. The forbidden products include pork, blood, or any animal not slaughtered in specific way. Similarly, *Islamic banking* provides banking services in accordance with Islamic Law, which enforces the sharing of profit and loss, and forbids the collection of interest[3]. Market for halal goods and services is growing fast, even outside Muslim countries, with, for instance, estimated sales of all types of halal food in Britain totaling £2.6bn in 2011[4]. However, new marketing approaches to Generation M extend its application beyond diet, emphasizing a whole *halal lifestyle*, which is maintained "not only in practice but also in philosophy" (Hussain, 2012).

The definition of this philosophy online encounters a crisis of *authority and authenticity* (Dawson and Cowan, 2013), with emerging "instant experts" existing outside traditional spheres of influence. This issue is especially acute in the context of halal, where quality, the process of manufacture and ethics, are central. The establishment of these halal qualities in the world market place in order to address the needs relating to purity, faith, health, etc. in turn leads to their commoditization in what Navaro-Yashin calls the "market for identities" (Navaro-Yashin et al., 2002; Wilson and Liu, 2010). As corporations such as Nestle and Unilever aggressively label their products as halal[5], it becomes increasingly important to understand the definition of halal by the new generation of consumers, and their relationship to old and new authorities on the matter.

Since its acquisition by Facebook in 2012, Instagram has become one of the most popular social media platforms in the world, with higher engagement than Facebook or Twitter[6]. It is used both by individuals and businesses to connect, express themselves, and maintain a web presence. To be searchable, the posts are often annotated with text and keywords (#hashtags), allowing for qualitative and quantitative analysis. Its unique scale and diversity makes it a rich resource for observational social studies of international scope. In this paper, we present a large-scale study of Instagram posts, representing a nearly complete set of posts up to April 2016 containing keyword "halal" (in English and in Arabic). We examine the meta-data associated with over 1 million English-halal and 300,000 Arabic-halal posts concerning their geographic distribution, languages involved, and perform content analysis in three major languages: Arabic, English, and Bahasa Indonesian. Driven by existing work (outlined below), we focus on the following research questions:

- What are the major **populations** around the world engaged in *halal* lifestyle and products?
- How strong is the connection to **religious authority** in matters surrounding *halal*?
- How strong is the association of *halal* with **healthy lifestyle**?
- What other sources of **authority** define *halal*, and do these authorities promote engagement?

To the best knowledge of the authors, this is the first large-scale study of halal culture on social media. Insights provided below contribute to the fields of internet, as well as Muslim, culture research, and to marketing professionals interested in this burgeoning market segment.

---

[2] http://www.fao.org/docrep/005/y2770e/y2770e08.htm
[3] http://www.investopedia.com/terms/i/islamicbanking.asp
[4] https://www.theguardian.com/lifeandstyle/2013/sep/25/halal-food-why-hard-find-britain
[5] https://www.bloomberg.com/news/articles/2009-10-06/germany-wakes-up-to-halal-muslim-food
[6] http://www.techradar.com/news/internet/the-reasons-why-instagram-will-be-the-next-big-platform-1306343

## 2 RELATED WORK

Since the popularization of social media, scholarly work has been ongoing in capturing the evolving cultural trends through this medium. Such studies span the American Black culture (Stokes, 2007), feminism (Schuster, 2013), generation gaps (Aksoy et al., 2013), and sexuality (Hasinoff, 2012). Online communities provide new affordances for expression (Leavitt, 2015), giving a fine control over the information users choose to share about themselves, allowing for a discussion of personal and controversial nature behind the screen of anonymity. Further, it has been shown that the representation of self in online environments then may feed back to the individual's self-perception (Yee and Bailenson, 2007). Thus, an ongoing dynamic re-definition of cultural norms is taking place. For instance, (Hadgu et al., 2013) find the political hashtags used on Twitter by one political party may be appropriated or hijacked by another, forcing their own meaning on an already popular search term. Another recent study on the use of hashtag #foodporn has revealed a subculture of fitness enthusiasts, imbuing the term with health-related slant (Mejova et al., 2016). Religious notions have likewise been affected.

*The Internet is changing the face of religion worldwide*, thus Lorne L. Dawson and Douglas E. Cowan begin their exposé in Religion Online (Dawson and Cowan, 2013). As religious symbols, notions, and practices find their expression online, they also reveal their evolving cultural context. In particular, several central research areas emerged around social practices, online-offline connections, community, identity, and authority online (Campbell, 2013). For example, a recent study of Islamic State of Iraq and Syria (ISIS) supporters examined the antecedents of ISIS support in their Twitter feeds, finding more references to Arab Spring uprisings which have failed (Magdy et al., 2016). Another examines the understanding of online privacy in the Persian Gulf, distinguishing it from the prevalent Western definition, emphasizing the gendered and religious aspects not implemented in most social media platforms (Abokhodair et al., 2016).

The Islamic notion of "halal" – that which is *permissible* – has been an especially rich topic, as it concerns religious identity, authority, and authenticity, and reflects the daily activities of thousands of Muslims (and as we shall see, potentially non-Muslims) online. Comprising more than a billion of the world's population, Muslims represent an important market segment – one which unites under the practice of halal – yet it is largely considered marginalized in Western mainstream marketing (El-Bassiouny, 2014). According to John L. Esposito and Dalia Mogahed, the central role of faith within Muslim migrant communities allows them to maintain a distinct cultural identity, while nonetheless adopting Western values such as democracy and human rights (Esposito and Mogahed, 2007). This then allows a development of a diversity of local Muslim sub-cultures around the world. Recently, social media has become a connecting medium for individuals and communities to communicate and construct their identity (Sandikci and Ger, 2010).

Recent small-scale studies have shed some light on the pressing issues of the halal culture online. A netnography by (Kamarulzaman et al., 2014) of six websites, including The Islamic Food and Nutrition Council of America (INFANCA) and yelp.com, found three major barriers to accessing halal, mainly supply sources, questions of authenticity, and questions of quality, which in part could be resolved via certification and religious authorities. Further, a questionnaire by (Mathew et al., 2014) has found that non-Muslims prefer halal due to food safety concerns, indicating a perception of it being more hygienic and of higher quality. Consumers may also view halal as exotic or interesting, even if they do not belong to the Muslim religion (Alserhan, 2010). Drawing on small-scale ethnographic analyses, these studies revealed an evolving and diverse use of halal on the web. Thus, *halal* in today's world is a multi-faceted concept. As (Wilson, 2012) asks about Islamic marketing, "Are we talking about a religion, a culture, an approach, or a business and management function?" In this paper, we provide a view of the worldwide halal culture via an approach grounded in large-scale social media analysis.

## 3 DATA COLLECTION

We begin by collecting two datasets using the Instagram Application Programming Interface (API)[7]. Instagram is a media sharing service in which a user can share an image or a short video, with accompanying description and other information, such where the picture was taken. Other users may then view, like and comment on these posts. Some users may restrict who sees their posts, and the posts of such users cannot be obtained via the API, and are not included in our dataset. In particular, we use the Media Search endpoint to collect two datasets, one using keyword *halal* in English transliterated spelling, and one in original Arabic, to create Halal English (shorthand, HE) and Halal Arabic (HA) sets, respectively. The Halal English dataset contains 1,004,445 posts belonging to 120,943 unique users and originating from 138 countries, while the Halal Arabic dataset contains 325,665 posts belonging to 11,516 unique users and originating from 66 countries.

The data gathering was conducted in April 2016 and consists of an exhaustive collection of all posts contained the keyword, resulting in data spanning four and five years, for HE and HA, respectively. The span of several years was intentional, to limit the effect of seasonal variation in posting and content behavior. The meta-data collected includes link to the media (image or video) posted, user name of the poster, location (if any), description written by the user, and the number of comments and likes. The media itself was not downloaded due to size and timing constraint. The analysis of the images and videos themselves poses an interesting future research direction.

## 4 GEOGRAPHY OF HALAL

We first examine the geo-coordinate meta-data, and find a relatively low percentage of located posts – 9.5% in HE and 5% in HA. Out of the located English posts, 54% are originating from South-Eastern Asian countries including Indonesia, Malaysia and Singapore. Figure 1 compares the number of halal-related posts from a country to the number of Muslims living in it (Hackett and Grim, 2012), normalized by the country's internet penetration obtained from the World Bank[8]. We find Malaysia and Indonesia at the top right corner, followed by Middle-Eastern countries, as well as major European ones of Great Britain, France, and Germany, as well as the United States. Pearson correlation between these is $0.37$ ($p < 0.001$), and Spearman rank of $0.49$, suggesting that the volume is quite representative of the overall volume of Muslim population throughout the world. One possible source of noise in this relationship is the uneven adaptation of Instagram in these diasporas across the world, however the company is reluctant to share detailed penetration statistics.

The top countries of Arabic data are from the Persian Gulf region, with Kuwait, Saudi Arabia, United Arab Emirates and Bahrain accounting for 89% of all geo-located posts. We omit a plot with Muslim population, as this dataset is extremely skewed, and accounts for fewer countries, not having a statistically significant Pearson correlation between the population and the captured Instagram posts (at $0.02$). However note that although most geo-located posts originated from Kuwait, this may not be the case for the rest of the Arabic data.

To provide a more comprehensive view of the cultural belonging of the captured posts, we apply *langdetect* package for Python[9] to the posting user's description text of each post. As expected, majority of documents in HA were identified as Arabic (95%), followed by Farsi (2%). However, HE data presented a more diverse set of languages, with 53% Bahasa Indonesian, 24% English, and a multitude of languages following, including German, Russian, and French. To ease the analysis of this heterogeneous data, we sub-divide it into Indonesian (HEI) and English (HEE), capturing two of the largest present languages. The distribution of the number of posts per user in each subset is shown in Figure 2. We find most Indonesian-speaking users posting more than the other two, however the mean posting rate is skewed heavily by power-users in Arabic dataset. As we later find, this is due to heavy commercial activity by the two groups. Note that we retain both the super-active users as well as singletons in the data for further analysis.

---

[7] https://www.instagram.com/developer/
[8] http://data.worldbank.org/indicator/IT.NET.USER.P2
[9] https://pypi.python.org/pypi/langdetect

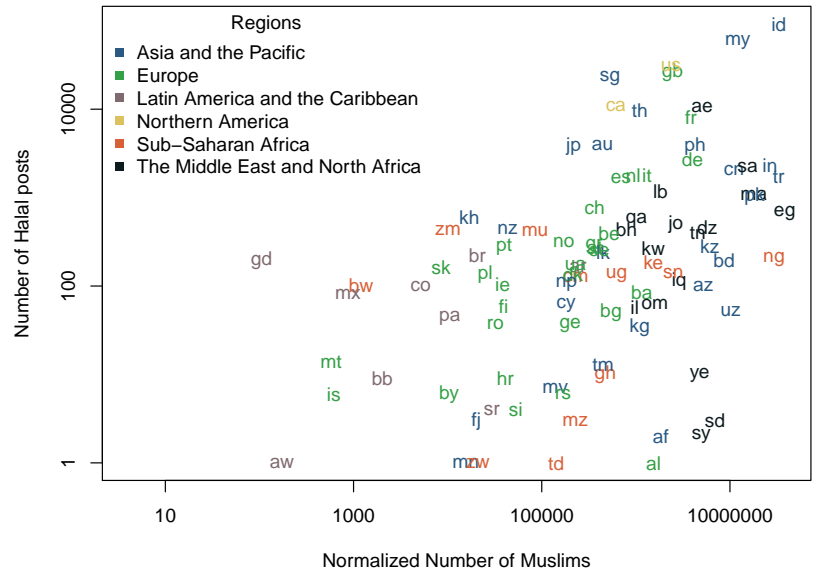

**Figure 1.** English data comparison between the number of halal-related Instagram posts captured in a country and the number of Muslims normalized by internet population, colored by major part of the world.

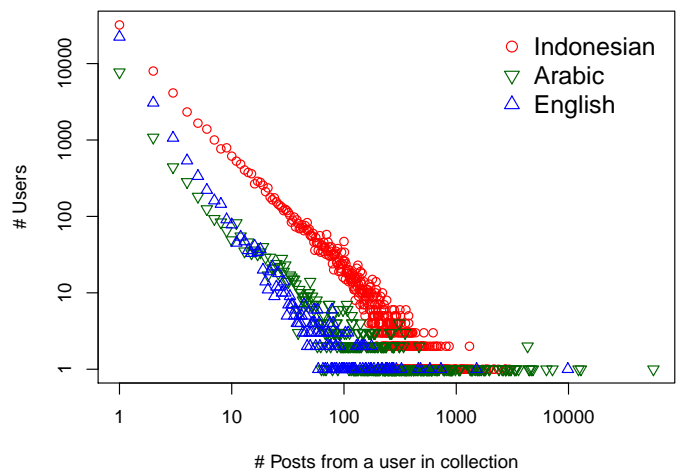

**Figure 2.** Distribution of posts per user in Arabic (green), and Indonesian (red) and English (blue) subsets of the English stream.

To extend the reach of geo-location, we "propagate" the countries by user, assigning the country of one geo-located post to all others for the same user (breaking ties randomly). Figure 3 shows the geographical distribution of the posts in Arabic, English (HEE) and Indonesian (HEI) datasets. For clarity, we do not normalize these figures by country's population, thus more populous countries have more chance to have more posts. Unsurprisingly, we find most Arabic post coming from the Middle Eastern region, and English

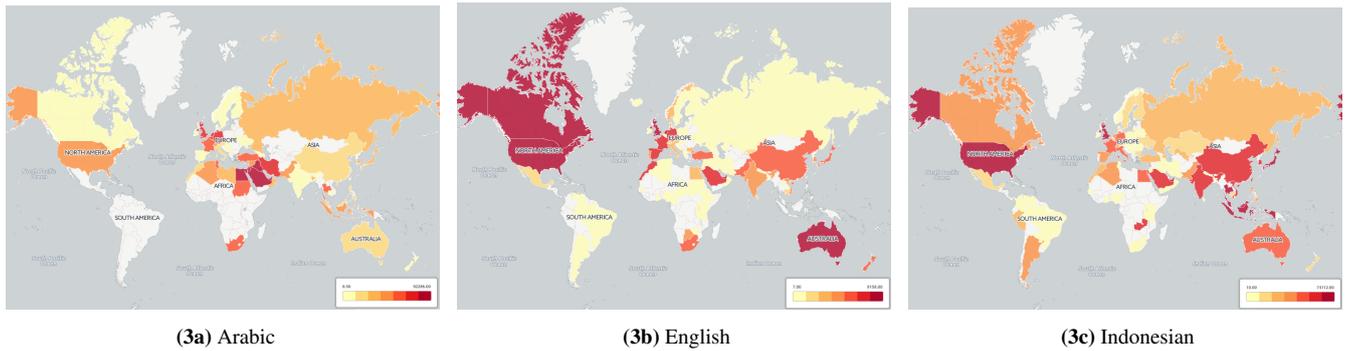

**(3a)** Arabic  **(3b)** English  **(3c)** Indonesian

**Figure 3.** Geographic distribution of the number of posts (un-normalized by population) for (a) Arabic, (b) English and (c) Indonesian language datasets.

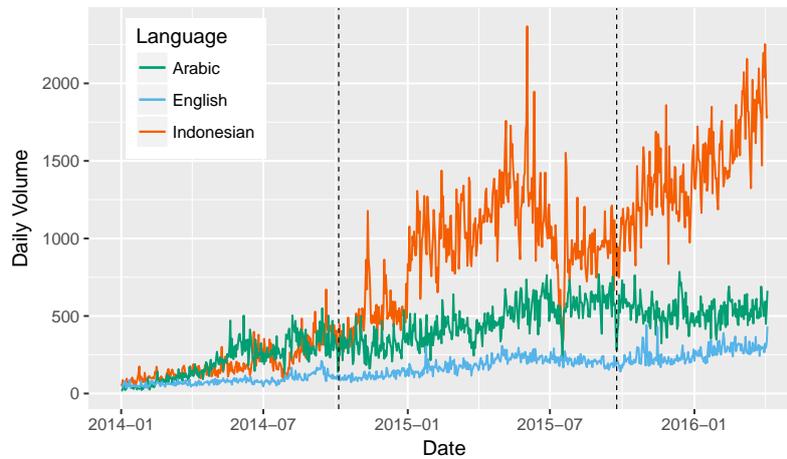

**Figure 4.** Temporal volume distribution for the Indonesian (blue), English (green) and Arabic (red) aggregated daily. The vertical lines signify the yearly holiday of Eid al-Adha.

from Canada, United States, Australia, and Western Europe. Interestingly, Indonesian data centers not only in Asia, but also United States, and many other countries, which is a testimony to the importance of Muslim community in Asia to the definition of *halal* on social media.

Finally, Figure 4 shows the temporal distribution of the data. It illustrates the increasing popularity of Instagram in all of the languages. We also show (vertical black lines) the annual holiday of Eid al-Adha, for which it is customary to purchase and consume halal sheep (note that we do not find a significant spike around that time for any language).

After the above dataset enrichment procedures we find three large populations using halal in their social media: Arabic speakers in the Middle East and Gulf, English speakers largely in the Western world, and Indonesian speakers of South Asia. Although sharing some cultural commonalities, below we show the great diversity in their use of *halal*.

## 5 GROUNDED CONTENT ANALYSIS

To explore the themes in the large amounts of data captured in the previous section, we begin with an automated topic discovery technique. We use Latent Dirichlet Allocation (David et al., 2003), a generative statistical model, to extract a number of topics from each collection. Intuitively, each extracted topic captures words which appear in similar contexts, resulting in (ideally) cohesive ideas that span many documents (in our case, posts). We apply JibbLDA[10], a Java implementation of LDA that uses Gibbs Sampling for parameter estimation and inference, to each of our three datasets. We experiment with several number of topics (often referred to as $k$), settling on 50 as resulting in most semantically cohesive output.

We then examine these topics using a hybrid approach. First, we annotate the topics utilizing Grounded Theory (Glaser and Strauss, 1999), which aims to discover themes in the data through repeated examination. However, we also pay a special attention to the elements of the data which are pertinent to themes discussed in the Related Work above. Thus, we begin by manually coding the 50 topics, examining the top 100 words of each, assigning primary codes to capture major themes, and secondary to record the thematic peculiarities. A separate set of topics for English, Arabic, and Indonesian has been analyzed by an author of this paper, each having native proficiency in either English, Arabic, or bahasa Indonesian. The process was done iteratively, with each successive version of the codes discussed between the authors, and with at least three passes made through the data.

Figures 5(a-c) show the co-occurrence networks of the resulting codes for each dataset. The edges (lines) signify the extent to which the two codes occur in the same topic, and the nodes (the codes themselves) are positioned using Force Atlas layout (Bastian et al., 2009), putting more central nodes in the center. The primary codes (in red) often occupy a central location in all of these, however the emphasis differs in each data. Also note that, despite the initial language of the data, the codes are presented in English, for the reader's convenience.

Arabic topics are often about *sheep breeds* and *trading*, with *food* and *fashion* being peripheral. Note that *Islam* is quite central and is associated with *scholars* and *Ramadan*, as well as with the central topics of trade. Communications keywords, including *whatsapp*, *Instagram*, and *blackberry* often appear alongside advertisements, indicating an informal market taking place through channels which are private and direct. United Arab Emirates (UAE), Kingdom of Saudi Arabia (KSA), and Kuwait play a central role in this trading, whereas food is often associated with Bahrain, which has a popular food delivery website *talabat.com*. Note two peripheral businesses of *fashion* from the Maghreb region (western North/Northwest Africa, west of Egypt) and of *banking* more closely associated with Kuwait and Qatar. Thus, we observe a conversation dominated by the animal trading, much of it through more private electronic channels, as well as peripheral association with fashion and banking.

The English-language conversation (note this data has been separated from the large Indonesian stream, as discussed in previous section), shown in Figure 5(b), revolves squarely around *food*. Its association with various health-related codes is extremely strong, including *vegan*, *organic*, *gluten-free*, *keto* and *weight loss*, as well as a strong presence of *supplements*. Here, we notice strong association with *FDA*, the US Food and Drug Administration, which provides certification for meats and other products[11] (note that no analogous authority was found in topics of the Arabic data). Also note the peripheral location of religious and Muslim keywords, with *Muslim* sometimes associated with *Kosher* (the Jewish dietary law), once again emphasizing the general "goodness" of the products, instead of conformation to a particular dietary religious law. This is further shown by the appearance of alcoholic beverages like *wine* in the topics.

Indonesian dataset shows a further removal from the world of food, with *supplements*, *skin care* and other cosmetic procedures being central to the conversation. In fact, keyword *whitening*, although not a primary code, takes the center stage of the network in Figure 5(c). Here, certification takes an even more central role: besides *FDA* we find *BPOM* (Badan Pengawas Obat dan Makanan is national agency of drugs and food control in Indonesia[12]), *MUI* (certification that is issued by Majelis Ulama Indonesia[13]), and *TGA* (Therapeutic Goods Administration, the regulatory body for therapeutic goods including medicines, medical devices, gene technology, and blood products in Australia[14]). Notice that the religious codes, including *Quran* and *Allah* are altogether separate from the main cluster, and the *health* aspect of these products is also peripheral. Finally, similarly to the Arabic data, we find an *Islamic investment* cluster, also separate from the main commerce-related conversation.

---

[10] http://jgibblda.sourceforge.net/
[11] http://www.fda.gov/AboutFDA/Transparency/Basics/ucm194879.htm

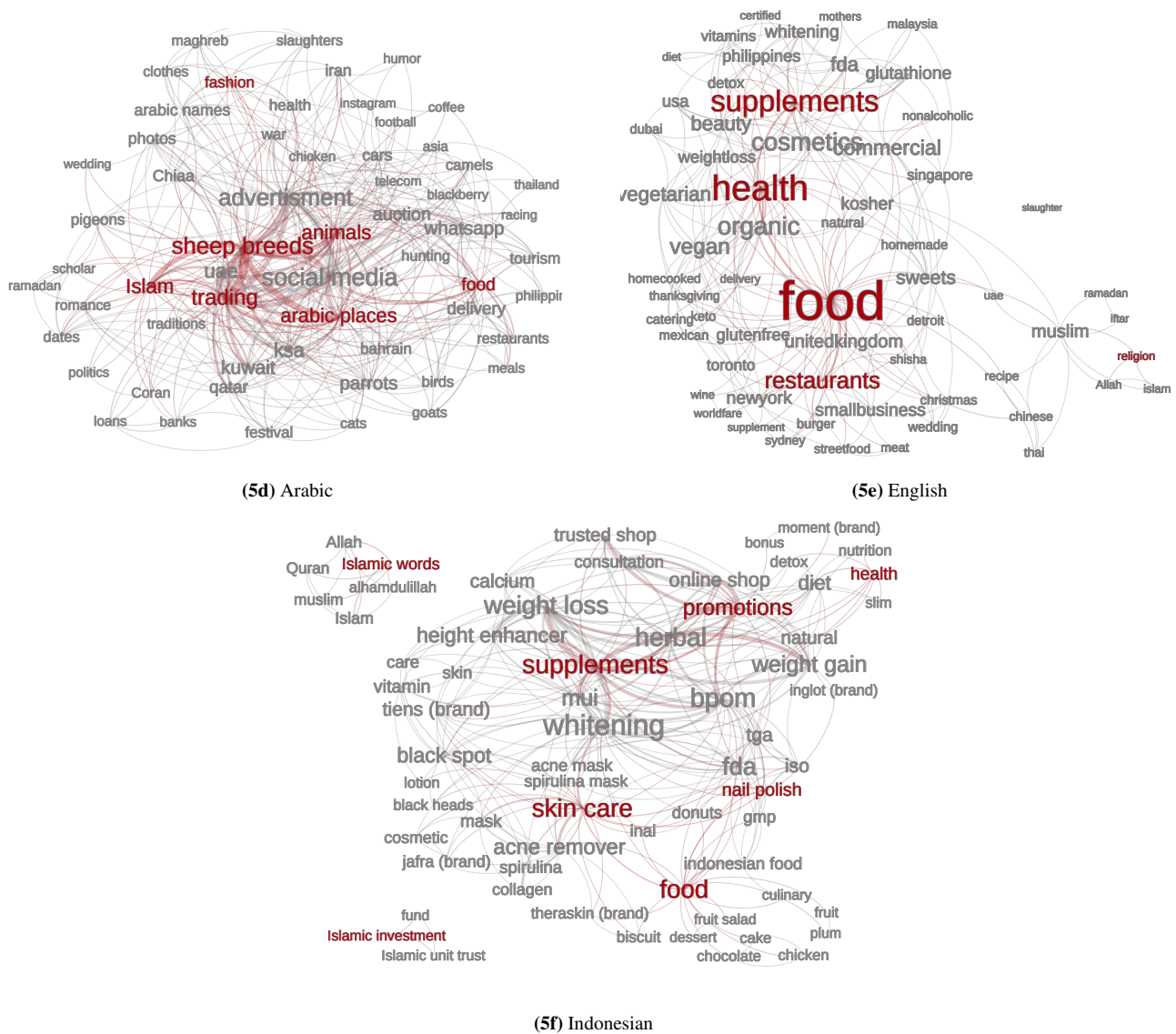

**Figure 5.** Code co-occurrence networks for the topics extracted from (a) Arabic, (b) English, and (c) Indonesian-language posts. Red codes are primary, gray are secondary codes, and the code placement reflects the centrality of the concept within the network.

---

[12] http://www.pom.go.id/new/
[13] http://www.halalmui.org
[14] https://www.tga.gov.au/tga-basics

Table 1. Statistics of language-specific lexicons, including the number of terms, the number of posts with at least one of these terms, and percentage of all data these posts represent.

| Lexicon | # words | # posts | % posts |
|---|---|---|---|
| *Arabic* | | | |
| Sheep/Animals | 164 | 198,795 | 61.0% |
| Trading | 58 | 84,411 | 25.9% |
| Islam | 56 | 41,671 | 12.8% |
| Food | 81 | 20,429 | 6.2% |
| *English* | | | |
| Food | 327 | 65,536 | 80.0% |
| Health | 123 | 36,397 | 44.4% |
| Supplements | 91 | 22,095 | 26.9% |
| Religion | 29 | 10,890 | 13.2% |
| *Indonesian* | | | |
| Food | 149 | 382,548 | 57.1% |
| Health | 63 | 461,892 | 69.0% |
| Supplements | 86 | 467,431 | 69.8% |
| Religion | 45 | 109,948 | 16.4% |

To summarize, we find distinct emphases in these three data streams. Arabic is largely about the trading of sheep and other animals, English about food, and Indonesian about beauty products and supplements. Each have a different connection to authority – religious (with Arabic having the closest ties), and institutionalized (Indonesian mentioning the greatest number of certification agencies). The association with health is also quite diverse, being central in the case of English-language data, peripheral in Indonesian, and almost non-existent in Arabic.

## 6 SOCIAL ENGAGEMENT

Next, we turn to the interaction of users with the posts, utilizing the above coding exercise to find the posts contributing to the themes identified above. To this end, we create custom language-specific lexicons for each dataset, covering major topics discovered in each data. These lexicons are described in Table 1, along with the number and percentage of posts in which at least one of the lexicon words appears[15]. Note that we only use high-precision keywords, such that the terms are unambiguously associated with the topic. For instance, English health lexicon may contain *diet*, *fitspo*, or *gym*, but not *good* or *beautiful* (or, trivially, *halal*). Thus, the (exact word string) matching which resulted in the shown number of posts is precision-oriented (i.e., conservative).

The Instagram API provides the number of likes for each post, and it is these markers of social interaction and approval which we examine in Figures 6(a-c). Note that the y-axis is on log scale, thus any visible change between the distributions is often substantial in practice. Overall, Arabic posts have on average more likes than English or Indonesian ones. The below observations are tested using one-way ANOVA F-test (Lomax and Hahs-Vaughn, 2007) and Welch two-sample t-test.

---

[15] *The lexicons are available at:* `https://sites.google.com/site/yelenamejova/resources#halal`.

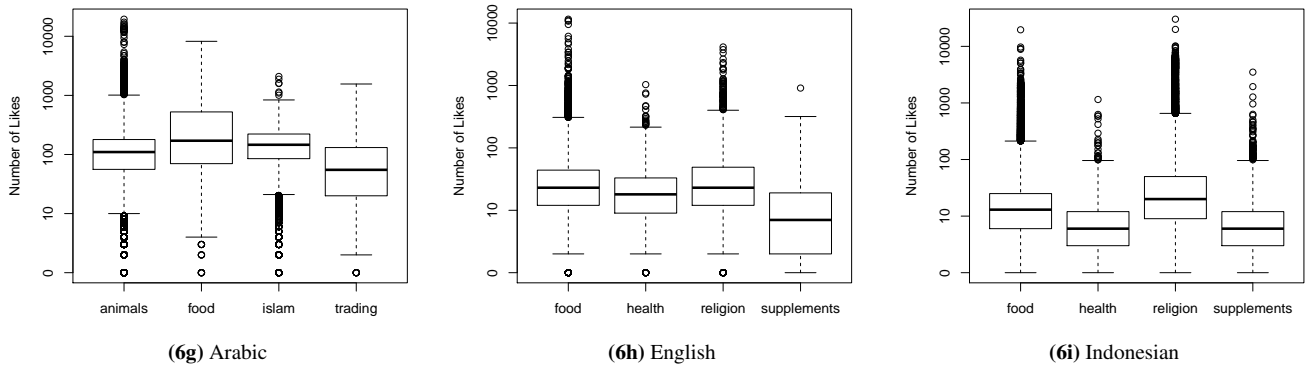

**Figure 6.** Distribution of likes in posts having at least one word matching a lexicon category, in datasets of the three languages.

Although most prolific category in Arabic data concerns sheep and other animals, the most liked topic contains words relating to food and restaurants, with on average 289 likes per post, compared to 168 of posts containing Islamic terms, and 145 animal-related ones. Trading has the fewest likes, on average 87. The difference between each of these means is significant at $p < 0.001$ level. This illustrates a potential mismatch between the supply of trade-related content, and the demand of the Arabic-speaking Instagram users, which we see in the other languages as well.

The English and Indonesian data share the same lexicon themes (recall that the terms wherein are different, specific to each language). Although the mentions of food dominate the English dataset, it is the content mentioning religious terms which has the higher average number of likes at 61, compared to food terms at 44 (health and supplements are at 33 and 22, respectively). Interestingly, posts containing food- and religion-related posts have the same median of 23 likes, indicating more abundantly liked content in the tail in religious content (the difference in means is still significant at $p < 0.001$ level).

A similar pattern can be observed in Indonesian data, where posts containing religious terms have an impressive 127 likes on average, compared to second most liked category of food at 34 (again, health and supplements at tail end with both 11 likes each). This difference is clearly visible in Figure 6 (and, again, is significant at $p < 0.001$).

Above observations can also be made by looking at the number of comments, however due to similarity of results, and the sparsity of comment information in our data, we omit these results.

Further, we look explicitly at the liking of the posts containing certification claims. In English, we focus on FDA, and in Indonesian also on agencies described above, including TGA and MUI. When considering posts mentioning food, in both languages the likes markedly plummet from 47 to 5 in English and 29 to 10 in Indonesian between posts not mentioning certification, and those that do. A linear model which incorporates the number of words in the post (which is positively related to number of likes), shows large negative coefficients for certification-related terms (see Table 2). These findings illustrate the lack of social engagement established governmental sources of certification garner online, with serious implications in the changes of authority in online setting.

Food posts with and without health-related keywords show the same behavior in both languages, with those *without* them getting many more likes, with similar implications for public health officials and private businesses interested in the association of halal with the healthy lifestyle.

**Table 2.** Linear models for English and Indonesian posts mentioning food, which use the number of words in the post text and whether a certification authority is mentioned to model the outcome variable of the number of likes. Estimated coefficients and their *p*-value levels are shown.

| **English** ($n$=65K) | | **Indonesian** ($n$=382K) | |
|---|---|---|---|
| Variable | $\beta$ | Variable | $\beta$ |
| (intercept) | 36.66*** | (intercept) | 29.10*** |
| $n_{words}$ | 0.17*** | $n_{words}$ | -0.06*** |
| $lex_{authority}$ | -63.79*** | $lex_{authority}$ | -2.16*** |

Significance: $p < 0.0001$ ***, $p < 0.001$ **, $p < 0.01$ *

The relationship between food and religion is more complex, however. When food-related posts are positioned in the context of religious keywords, in the case of English, the average likes increase (significantly) from 36 to 45 (median of 16 to 26). In the case of Indonesian data, the increase of average from 16 to 20 (sig. $p < 0.001$) corresponds to a decrease in median from 9 to 8, indicating a substantial tail at play. Finally, in Arabic, religious keywords decrease the average likes from 168 to 144 (sig. $p < 0.001$). Once again, we find the shifting social engagement with religious authority, which we discuss below.

To further understand the nature of the popular content which may be swaying these numbers, we turn to the popular users of our datasets.

## 7 USERS AROUND HALAL

Next, we examine the users who are at the forefront of defining "halal" on Instagram. We define these in two ways: users who are most prolific (posting the greatest number of posts), and those whose content has engaged its audience the most (having the greatest average likes of their halal posts). Related to the second, we also look at users who are the most "lifetime" liked, that is, those who over time have gathered the most likes.

The most prolific users, in all three languages, are the commercial entities selling sheep and animals in the case of Arabic, food and supplements in English and Indonesian. These accounts number in tens of thousands of posts, and often have a range of products beyond halal food, spanning cars and nail polish, and almost never mentioning religious texts. For example, an account from United Arab Emirates has contributed 57,065 posts to the Arabic dataset, with each post on average having 230 likes. However not all businesses are so popular: the most prolific account in the English dataset focuses on dietary supplements. Its 39,878 posts on average get only 4 likes. Thus, by volume, these retailers are taking advantage of the easy access to social media for marketing purposes, but may have different success at engaging their audience.

However, the likes bring out altogether different users, with the most liked posts generated by celebrities. The top ten Arabic accounts include humorous accounts, and those of actors and political commentators. One popular account promotes Shia Islam and the Iranian government, associating halal with a political ideology. English and Indonesian accounts receiving the most social attention are often from entertainment industry – musicians, models, and actors. We find the definition of halal beyond food, as a part of an identity. Throughout the data, we find examples of posts which received thousands of likes, for instance promoting a "halal" fashion show, "halal" fitness culture, a "halal" movie (also tagged with *#legalonlinemovie*), and other lifestyle posts emphasizing clothes and stylish surroundings. The visual appeal of the content is paramount. For instance, the user garnering the most likes on all of their posts in our dataset is a visual artist, posting daily quotes from Quran and other motivational content, each of which receives thousands of likes. However, food remains a popular topic, as many other top "lifetime" liked users are food delivery businesses, posting attractive images and videos of their selections, as well as recipes.

## 8 DISCUSSION

Instagram is one of the most popular social media platforms, having hundreds of millions of users, and benefiting from its association with Facebook[16]. The images shared there encompass both private and business accounts, giving us a perspective on both sides of the halal market. Social media users, though, are notoriously young, and from more affluent walks of life – precisely the *Generation M* Shelina Janmohamed defines (Janmohamed, 2016). Thus, results we present here apply especially to this cohort, and may under-represent the opinions of older generations, those not actively using social media, and businesses not having a web presence. However, the growing adoption of mobile technology and social media use is continuing to enlarge the populations social media studies such as ours are able to capture. For instance, in 2016 the number of mobile social media users grew by 30%, up an impressive 581 million, particularly in South and Southeast Asia (We Are Social, 2017).

In this study, we have attempted to cover most large populations who may use halal keyword in their social media presence, but we surely missed the smaller diasporas who use different alphabets, as well as alternative spellings (such as using numbers for Arabic letters such as "7alal"). However, the dataset gathered in this study may be useful in bootstrapping additional keywords for future research. More geo-tagging efforts are also needed to further sub-divide the data into regional and national samples.

Whether or not the definition of "halal" by Islamic law is open to diverse interpretation to its adherents, its positioning within existing cultural spheres on Instagram varies widely across the world. Grouping our users by language (subsequently targeting different parts of the world), we find distinct usages of halal. In each language, though, commercial interests are the most aggressive in associating halal with their wares. Although both English and Arabic language posts speak about food, Arabic posts emphasize the live animals sold for slaughter, and less so restaurants. The increasing obesity epidemic in the Middle East is often attributed to the increasing prosperity of the region, allowing people to eat out (Alzaman and Ali, 2016). However, we show that the traditional animal traders are eagerly adopting social media marketing. Thus we would caution public health researchers from focusing on Westernization of the Middle Eastern diet at the cost of ignoring the health effects of traditional halal cuisine.

Further, we find a strong association of halal with healthy lifestyle in the English language posts, though some medical professionals claim halal (or, similarly, kosher) meat is no healthier than the regular kind (Park, 2015). There are those who maintain halal encourages humane animal treatment, and promotes food safety (Rezai et al., 2010). Regardless of the interplay between religious and governmental food safety standards, we find a strong connection between halal and healthy lifestyle. In particular, fitness products and supplements market themselves as halal, potentially reaching a new market segment. Health officials concerned with the improvement of the general health of Muslims may want to take similar approach.

Moreover, we find a complex relationship between mentions of religious keywords and popularity of the posts. Although secularization is thought to largely take place in the Western world (Marranci and Marranci, 2010), our results, on contrary, show fewer likes for posts with religious terms in Arabic, and not

---

[16] http://www.techradar.com/news/internet/the-reasons-why-instagram-will-be-the-next-big-platform-1306343

English or Indonesian. A further study of the demographics of these users will more clearly delineate the characteristics of Generation M, and the potential differences with the older generation.

An even stronger disassociation from authority is found in posts mentioning regulatory agencies such as Food and Drug Administration (FDA) in United States and Majelis Ulama Indonesia (MUI). Even when we control for the length of the post, the contexts in which these "authority" markers appear, or not, may differ substantially. For instance, it may be the case that celebrities never mention these, never exposing large audiences to these standards. However, if these authorities target social media users[17], quantitative analysis of the social success their content achieves may be an effective strategy for better targeting.

We note here that the comments captured in this dataset are not amenable to classic Sentiment Analysis, as "natural language" is foregone by Instagram users in favor of enumerating hashtags. Our experiments in applying state of the art sentiment lexicons showed a weak signal not related to any variables of interest in this work.

However, the media content – the images and videos – of the Instagram posts present a rich avenue for future research. This would involve the deployment of image processing algorithms, such as those detecting people (Mathias et al., 2014), food (Yanai and Kawano, 2014) and health information (Weber and Mejova, 2016; Kocabey et al., 2017), as well as cross-referencing this data with the available metadata and user annotations. As long as the necessary semantic information can be extracted from the images, possibly with the supervision of subject matter experts, it may greatly enrich our understanding of cultural expression on social media.

Finally, an important consideration of social media studies is the preservation of privacy. Unlike Twitter, but similar to Facebook, Instagram users can set their account to private mode, so that only approved users can view one's posts and other activity[18]. Similarly, such private media is not returned by the Search API. It also means that dataset described in this work does not captured the halal context in private conversations, which may differ from the public sphere. Future work in more personal understandings of halal and other religious matters may include anonymous surveys and face-to-face in-situ interviews.

## 9 CONCLUSION

In this paper we describe the first large-scale study of *halal* on social media. Retrieving over a million posts on Instagram mentioning halal in English and Arabic alphabets, we capture several distinct populations around the world. We find a surprising diversity in halal contexts around the world, where the mentions of authoritative topics, both concerning religion and government institutions, do not always garner more social interaction. We hope these insights encourage Muslim culture scholars, as well as those interested in marketing to the new generation of Muslims online, to use data-driven techniques in discovering the new meanings of halal which are developing on the social web. An exciting future research direction is applying automated image analysis techniques to the actual images shared on social media in order to enrich the scope of captured context, including social and environmental.

---

[17] These agencies do have a web presence, for example, MUI has a Twitter profile `https://twitter.com/Warta_MUI`
[18] https://help.instagram.com/116024195217477